# Time resolved evolution of the 3D nanoporous structure of sintered Ag by X-Ray nanotomography: role of the interface with a copper substrate.

*Kokouvi Happy N'Tsouaglo, Xavier Milhet\*, Jérôme Colin, Loic Signor, Azdine Nait-Ali, Juan Creus, Mikael Gueguen, Pascal Gadaud, Marc Legros*

K.H. N'Tsouaglo, Dr. X. Milhet, Pr. J. Colin, Dr. L. Signor, Dr. A. Nait-Ali, Dr. M. Gueguen, Dr. P. Gadaud

Institut Pprime - UPR CNRS 3346, Dept. Physics and Mechanics of Materials, ENSMA – Université de Poitiers, 1 avenue Clément Ader, 86961 Chasseneuil-Futuroscope, France

E-mail: xavier.milhet@ensma.fr

Pr. J. Creus,

LASIE – UMR CNRS 7356, Université de La Rochelle, Pôle Sciences et Technologie, Avenue Michel Crépeau, 17042 La Rochelle, France

Dr. M. Legros

CEMES – CNRS – 29 rue Jeanne Marvig 31055 Toulouse cedex 4, France

## Abstract

The evolution of the nanoporous structure of cylindrical sintered silver samples during high temperature aging, ranging from 200 °C to 350 °C for 350 min, is studied through in-situ Computed X-Ray Tomography. Investigations were performed for two types of specimens: pure sintered silver and specimens containing a silver-copper interface. It is shown that the overall pore evolution is driven by the evolution of very few large ones. The smaller pores, although being more numerous, do not really evolve before being absorbed by the few bigger ones. In pure silver, pore evolution is driven by diffusion (Ostwald ripening) but the presence of an interface promotes faster growth kinetics until the aging time reaches a threshold value, after which a deviation from Ostwald ripening occurs. The transition is a function of the aging temperature. This behavior is associated with the competition between elastic relaxation and surface energy minimization.

## 1. Introduction

The evolution of the pore structure in sintered Ag is critical for the sustainability and performance of power electronic modules. The most advanced power modules are developed around SiC or GaN chips, allowing better performance at the cost of higher operating temperatures. [1-4] The porous sintered joint used to bond the module to the heat sink is critical for the properties, and its microstructural evolution was reported several times in the literature using surface or two dimensions (2D) observations.[5-7] However, 2D images are relatively difficult to interpret, especially when dealing with the porous structure evolution during temperature aging. For sintered silver, pore evolution was reported to follow Ostwald ripening, for which a shift in average pore radius from 80 nm toward slightly larger values is observed. [5] Some authors report densification while others don't. [5-9] The discrepancy between those studies might come from a complex porous evolution, difficult to interpret using only 2D observations. Attempts to link 2D to 3D were performed using several sequential cuts within a SEM, either using Focused Ion Beam (FIB) or in-situ serial block face scanning electron microscopy.[10-13] These two techniques, that involve SEM imaging, possess the appropriate resolution to observe the nanometric pores found in sintered Ag. However, these methods are destructive and therefore, do not allow to follow the porous structure evolution during aging. On the other hand, resolution progresses in X-Ray computed tomography (XCT) have promoted a new, non-destructive approach either in 3D or in 4D, when in-situ techniques (e.g. evolution with time) are implemented within the beam line (see for example [14-27]). It has proved to be a powerful tool to monitor microstructure tranformations occurring during aging, solidification, deformation…[14-28]. In particular, 4D XCT brought some new insights about precipitates evolution since 3D imaging allows a more accurate description of the microstructure compared to standard 2D observations.[25-27] However, to the authors'best knowledge, data on the evolution of a nanoporous structure in 4D are scarce, only at temperatures barely exceeding 250 °C and for relatively short aging times (180 minutes max).[19] In these conditions, the porous structure of pure sintered Ag was reported to evolve according to Ostwald ripening and no densification was observed. However, power modules are complex engineered systems that involve several layers of different materials. As a result, many interfaces are found in the stack. All of these materials exhibit different properties such as, for example, different thermal expansion coefficients (CTE). In operating conditions, heat is generated by the chip and the CTE mismatch leads to the development of thermal stresses in the stack. To the authors' best knowledge, the potential influence of interfaces (e.g. a porous film deposited on a substrate for example) on the porous structure evolution of the film has been very seldom reported while this might be a critical issue for the reliability of power modules. [20, 29, 30] Compared to the state-free state, for which precipitates evolution is driven by diffusion (LSW theory and Ostwald ripening), elastic stresses may modify the coarsening behavior.[31-38] In stress-free materials, precipitates evolution is driven by surface energy minimization while in elastically stressed solids, this evolution is driven by a competition between elastic and surface energies. For instance, a change in the evolution of the precipitates shape was reported to occur during long term aging of a Ni-Ti-Al alloy. This was attributed to the loss of coherency (i.e. elastic stresses relaxation by nucleation of misfit dislocations).[38] Here, we study the evolution of the porous microstructure of either pure sintered silver specimen (s-Ag) or specimens including an interface (s-Ag/Cu) with temperature using 4D X-ray nanotomography allowing in-situ monitoring (time resolved 3D analysis) at a sub-micron scale. The latter configuration mimics what can be found in a realistic system and promotes the generation of thermal stresses in the stack. Evolution of the pores was closely monitored at 200 °C, 260 °C, 300 °C and 350 °C. Drastic differences in the evolution of the porous structure during temperature aging are observed between pure silver and specimens containing an interface with copper. An explanation of the structural evolution is provided in the framework of stress-driven theory where the interfaces and free-surfaces of stressed solids develop periodic fluctuations of well-defined wavelengths ($\lambda$) to release the stress.[39-43] In this context, the effect of thermal aging on the morphological evolution of the pores with or without stress is discussed.

## 2. Results

### 2.1. As-sintered microstructure

Figure 1 shows an example of the typical nanoporous microstructure of sintered silver. The pores can be divided in sub-groups based on their average radius. It must be emphasized at this step that 3D Tomography presents the advantage over 2D experiments to allow an accurate visualization of the pores, their shape and their connectivity.

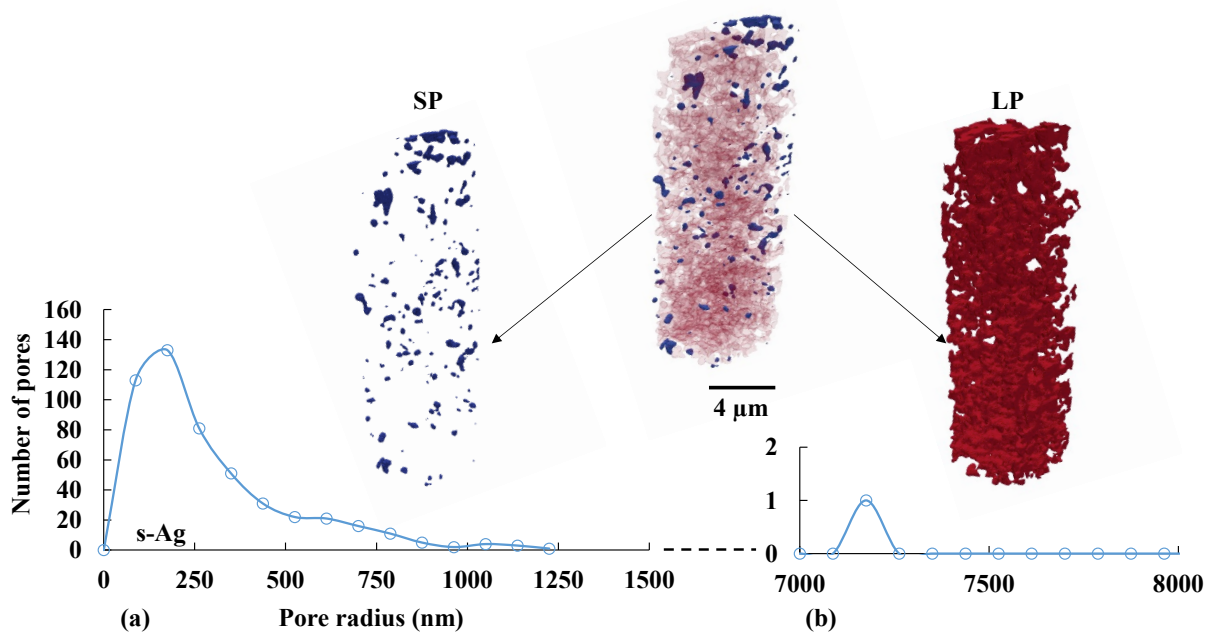

**Figure 1**. Pore distribution within the volume of an as-processed Ag pillar consisting of two groups based on their radius: a) Size distribution **of the numerous** isolated small pores (SP - in blue on the tomogram) vs b) **1 very large pore** (LP - in red)

**Figure 2** shows an example of the discrepancy between surface (2D) and volume (3D) analyzes of the same area: the (3D) volume (2 x 2 x 2 μm$^3$) contains only 13 pores (**Figure 2a**) while 92 «independent pores» can be counted in the (2D) surface, obtained by a random cut within this (3D) volume, (**Figure 2b**). In addition, 3D data allows one to measure the volume of each individual pore. Once the volume of a pore is known, a sphere with the same volume can be easily obtained. The apparent diameter of each pore derives from its own equivalent sphere. Interestingly, only two subgroups can be created for each of the tested specimens (**Figure 1a** and **Figure 1b**): the first group, named SP (for smaller pores), contains over 99% of the total number of pores (**Figure 1a**) and the second one, named LP (larger pores) contains only very few pores (generally in the range 1 to 3 maximum) (**Figure 1b**). It must be pointed out that the average diameter of the pores in LP is generally one order of magnitude larger than those found for the pores belonging to the SP group.

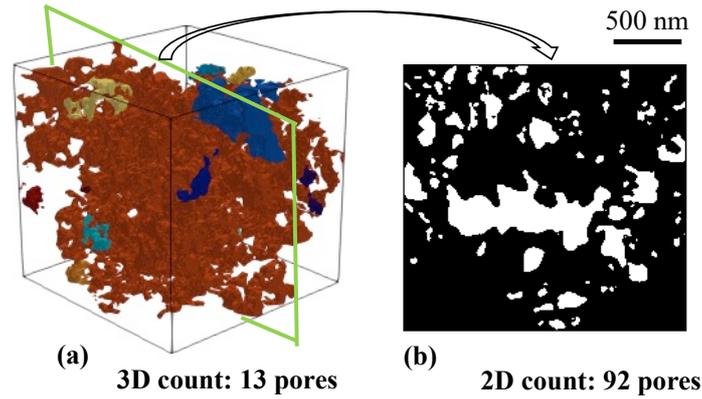

(a) 3D count: 13 pores  (b) 2D count: 92 pores

**Figure 2.** a) Example of a reconstructed 3D volume of s-Ag using Avizo vs b) surface analysis (obtained after a cut following the green plane defined on the 3D rendering)

**Table 1** reports the volume ($V_{LP}$) and number ratio ($N_{LP}$) of the larger pores against, respectively, the total pore volume ($V_T$) and total pore number ($N_T$) in the specimen. As observed in Table 1, the connectivity of the pore volume is massive in sintered Ag: few large pores (LP group) represent over 80% of the total pore volume while representing only around 1% of the total pore number. These results are totally in agreement with previous 3D studies reported in the literature.[18]

**Table 1.** Larger pore (LP) vs total porous volume $V_T$ and pore number $N_T$ (in percentage) for each Ag specimen. The subscript indicate pure Ag (s) or Ag/Cu pillar (c)

| Sample | 1s | 2c | 3s | 4c | 5s | 6s | 7c |
| --- | --- | --- | --- | --- | --- | --- | --- |
| $V_{LP}/V_T$ [%] | 80 | 91 | 94 | 91 | 84 | 88 | 61 |
| $N_{LP}/N_T$ [%] | 0.80 | 0.91 | **0.56** | 1.06 | 3.28 | 0.41 | 1.15 |

## 2.2. Density

**Figure** 3 shows an example of the evolution of the density for four specimens, where the dashed lines are added to guide the eyes. Density is obtained using the following method: i) first, the ratio $F_v$ between the volume of the pores and the total volume of the specimen is calculated for each Ag specimen. $F_v$ corresponds to the pore volume fraction within the specimen and ii) the density of sintered Ag is deduced using $F_v$ and Ag bulk density (10.5 g.cm$^{-3}$). As observed in **Figure 3**, after a slight evolution at the very beginning of the aging process, the density of the specimen remains relatively constant over time. This remains true even after aging up to 8h at very high temperature (350 °C) whether the specimen is pure Ag or Cu-coated. The slight evolution before stabilization takes a bit longer at 200 °C but is not significant. Overall, no densification is observed during aging. This was already reported either by surface analysis in 3D at temperatures up to 300 °C for as long as 3000 h.[2, 5, 8, 9, 16, 19] As a result, the pore evolution reported in the next section, occurs at constant pore volume fraction (within the normal experimental standard deviation ± 0.5 g.cm$^{-3}$).

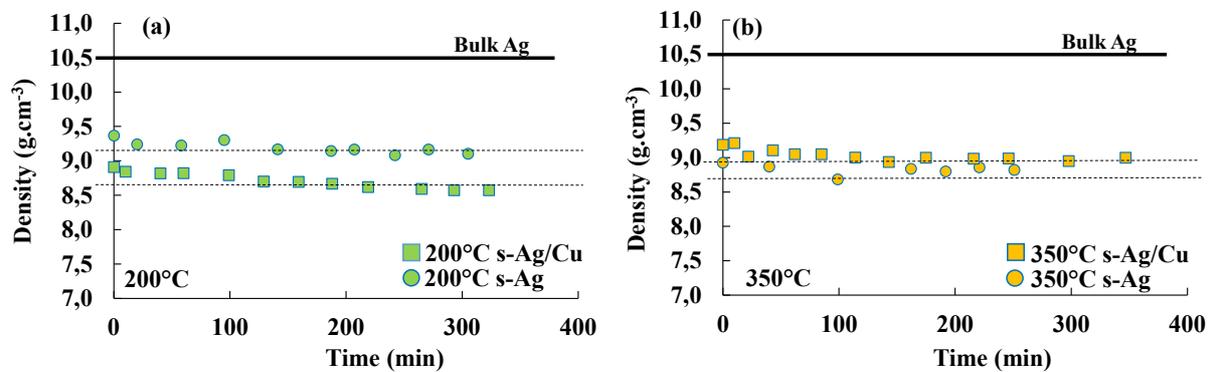

**Figure 3.** Evolution of the density vs aging time at a) 200 °C and b) 350 °C for s-Ag and s-Ag/Cu

## 2.3. Pore microstructure evolution vs aging time: role of the Ag/Cu interface

The evolution of the porous structure of pure sintered Ag (s-Ag) and sintered Ag with a Cu interface (s-Ag/Cu) specimens were carefully monitored during temperature aging. **Figure 4** presents qualitatively a typical example of the evolution of the porous structure of sintered Ag.

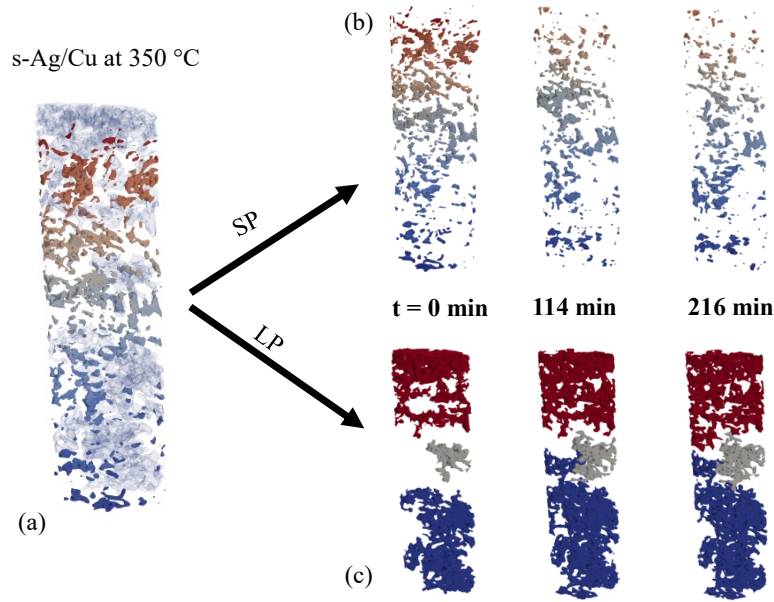

Figure 4: Example of evolution of the porous structure within a s-Ag/Cu specimen aged at 350 °C up to 216 min: a) entire porous structure, b) numerous small pores (SP) and c) 3 large pores (LP)

Obviously, each specimen is slightly different from each other (initial pore number, average pore radius). In order to give a comprehensive description of the microstructure evolution depending on the aging condition, all the data have been normalized using one of the specimens. **Figure 5** presents the evolution of the average pore diameter versus time for various temperatures: 350 °C (Figure 4a), 300 °C (Figure 4b), 260 °C (Figure 4c), and 200 °C (Figure 4d) for both s-Ag (except for 260 °C) and s-Ag/Cu using the classical Lifshift, Slyozov and Wagner ( LSW) time evolution law:[31, 32]

$$(R^3 - R_0^3) = K.t \qquad (1)$$

where R is the radius at a given time t, $R_0$, the initial particle radius and K, a rate constant for coarsening depending on the temperature T.

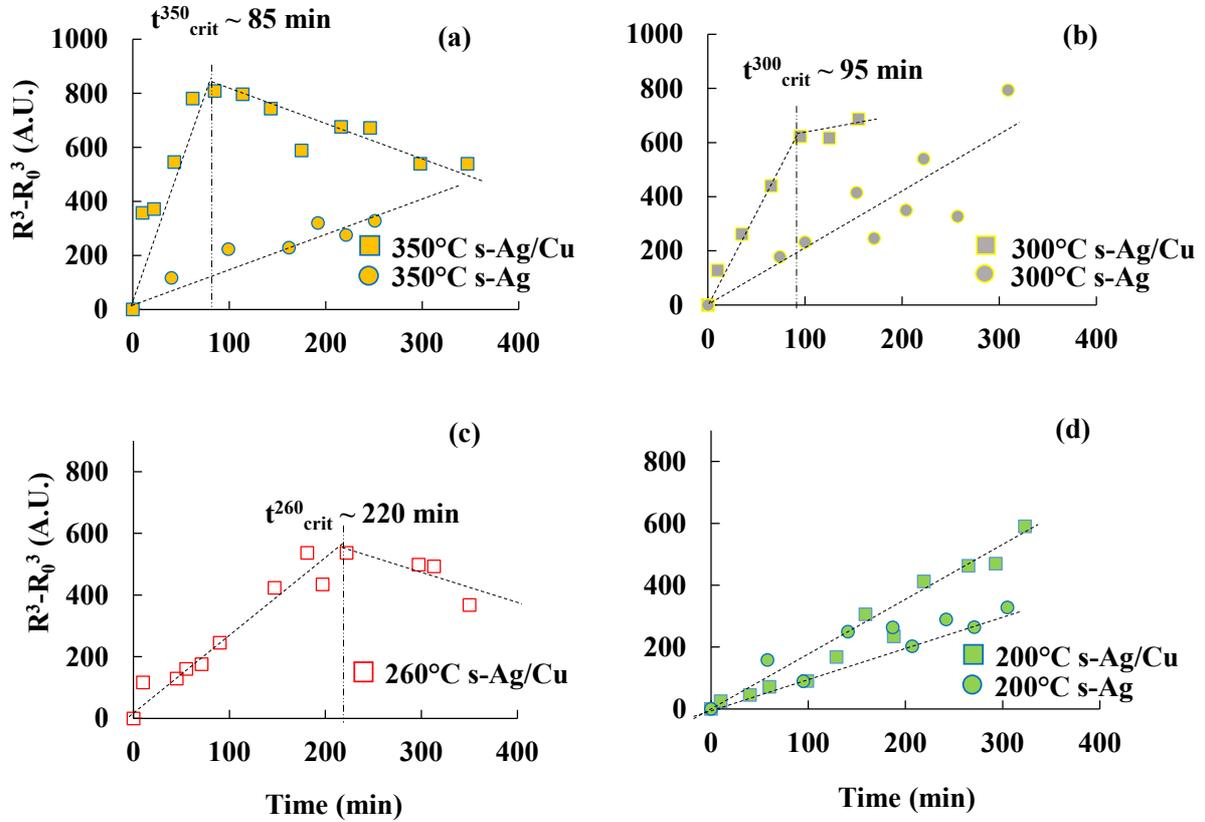

**Figure 5.** Average pore radius evolution (using a LSW representation [32]) vs aging time for S-Ag and s-Ag/Cu at a) 350 °C, b) 300 °C, c) 260 °C and d) 200 °C

In this LSW representation, classically used for Ostwald ripening studies, linear growth over time is assumed to be driven by diffusion.[30] In the case of s-Ag, the pore growth is linear, in agreement with Ostwald ripening. In the case of s-Ag/Cu, a deviation from the monotonic growth (Ostwald ripening) is observed after a critical time $t_{crit}$ except for the specimen aged at 200 °C. In this latter case, no deviation is observed. If only the linear initial stage (e.g. when t remains under $t_{crit}$) is taken into account for all the s-Ag/Cu, the growth rate increases compared to s-Ag aged at the same temperature, even for the specimen aged at 200°C. It can be pointed out that, as reported in earlier studies for this system, the higher the temperature, the faster the growth rate.[19] Actually, it can be noticed that the growth rate at 300°C is slightly higher than that at 350 °C for s-Ag. It would be expected that the higher the temperature, the faster the growth kinetics. As it can be noticed in Figure 5, the standard deviation for the 300 °C s-Ag/Cu data is much higher that those for any other data we obtained. $R^2$ is only of 0.76 for the data at 300 °C while $R^2 = 0.9$ for those at 350 °C. Therefore, it is difficult to extract a reliable tendency for 300 °C. The presence of an interface enhances this behavior (higher temperature / faster growth kinetic) before the deviation from Ostwald ripening occurs (**Figure 6**).

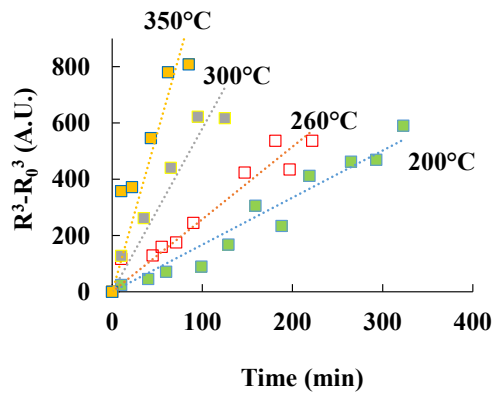

**Figure 6.** Influence of the temperature on the Ostwald ripening (before $t_{crit}$ – see Figure 5) for sintered silver pillars containing an Ag/Cu interface (s-Ag/Cu). For a given temperature, the observed slopes are always steeper than for pure Ag.

**Figure 7a** and **Figure 7b** show the evolution of the number of pores against aging time for s-Ag and s-Ag/Cu respectively. As expected, the number of pores decreases with aging time for s-Ag. In contrast, the number of pores increases in s-Ag/Cu once $t_{crit}$ is reached, except, once again, during aging at 200 °C, where only a decrease is observed. As expected, a higher temperature and an Ag/Cu interface helps for a faster decrease of the pore number (until $t_{crit}$ for s-Ag/Cu at 350 °C, 300 °C and 260 °C).

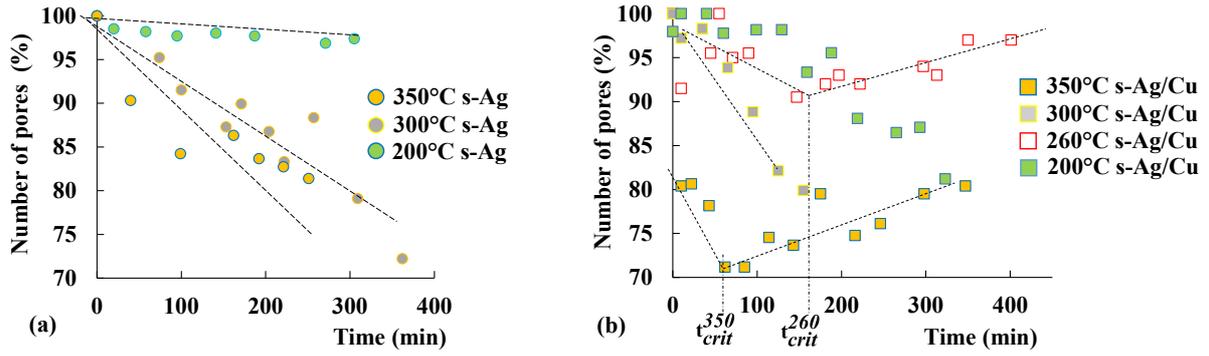

**Figure 7.** Evolution of the number of pores vs aging time for a) s-Ag and b) s-Ag/Cu. For the S-Ag/Cu, a transition occurs after t$_{crit}$

As mentioned earlier, the pores are distributed in two sub-groups, SP and LP, defined by their radii. **Figure 8a** and **Table 2** show a typical example of the evolution of a SP pore cluster. As observed in **Figure 8a**, the evolution during aging between pores belonging to SP is relatively surprising: in as-sintered condition, the area contains only one pore (Figure 8a – 0 min) with a sphericity of 0.41 (see Table 1 – 0 min). Here, the sphericity (Sy) corresponds to the ratio between the equivalent sphere surface (the equivalent sphere is defined by using the equivalent pore radius, deduced from the measurement of the equivalent volume of the pore $V_p$) and pore surface deduced from the experimental shape. During aging, the connectivity decreases as the pores split into three pores (**Figure 8a** – 10 min). At this point (after 10 minutes aging), the sphericity of pore gets close to 1 (**Table 1** – 10 min). When aging carries on, the pores reconnect with each other (i.e. the connectivity increases), promoting a decrease of the sphericity of the new entity **(Figure 8 and Table 1 – 40 min).** This complex evolution e.g. splitting/connecting repeats on and on during aging for the pores of the SP group.

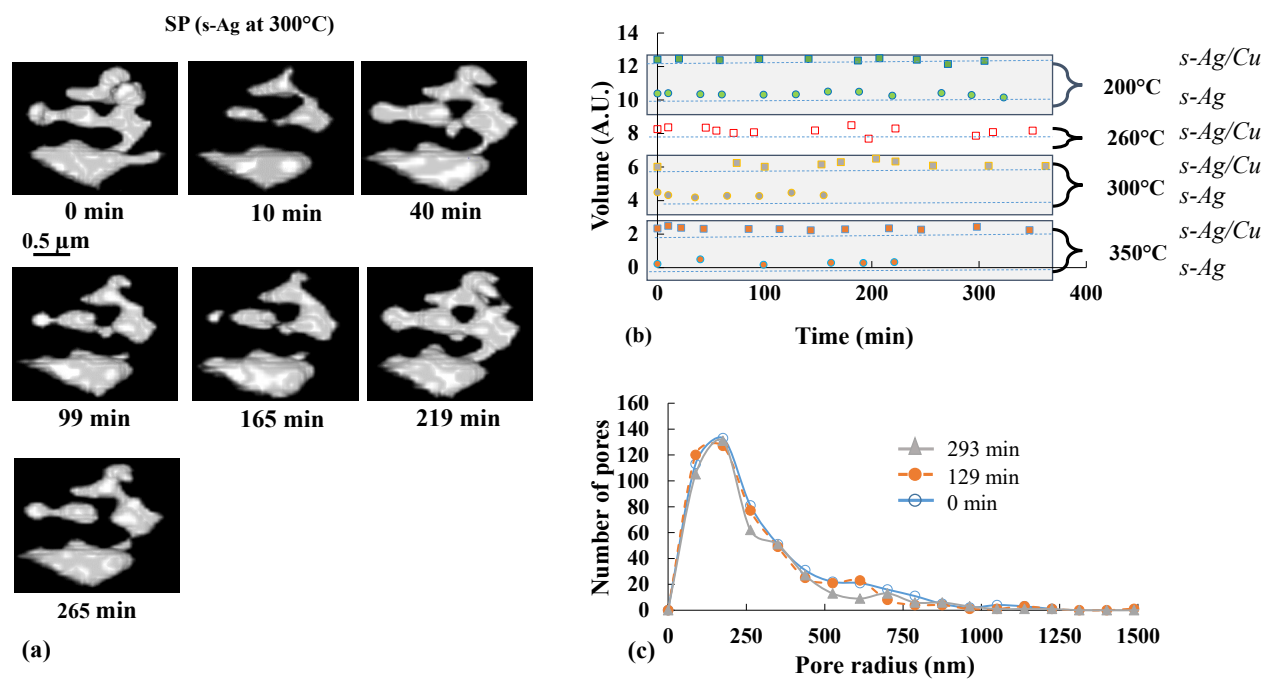

**Figure 8.** Example of the splitting/reconnecting evolution of pores from the SP group with aging time a) pore cluster, reconstructed from XCT, b) volume vs aging time and c) number of pores vs diameter for various aging time

**Table 2.** Example of the evolution of the volume ($V_T$) and sphericity (Sy) for a cluster of small pores (SP group)

| Time [min]   | 0    | 10   | 40   | 99   | 165  | 219  | 265  |
|---|---|---|---|---|---|---|---|
| $V_T$ [μm³] | 0.70 | 0.38 | 0.64 | 0.50 | 0.51 | 0.75 | 0.55 |
| Sy           | 0.41 | 0.60 | 0.45 | 0.66 | 0.60 | 0.44 | 0.50 |
|              | -    | 0.75 | -    | 0.51 | 0.67 | -    | 0.74 |
|              | -    | 0.81 | -    | -    | 0.82 | -    | -    |
|              | -    | -    | -    | -    | 0.93 | -    | -    |

**Figure 8b and Figure 8c** plot respectively the evolution of the average volume and pore diameters of SP pores during aging. As observed in these figures, the SP pores do not seem to evolve at all, their average volume being stable over time. Surprisingly, while the initial pore diameter is very similar between the two studies (3D and 2D), the evolution observed in **Figure 8c** differs from data obtained using surface analysis reported in a previous study, for which a shift toward larger radii was **systematically** observed.[5] This is likely the result of the existing discrepancy between 2D and 3D analysis: 2D cannot account for any potential connectivity within the volume (**Figure 2 and Figure 8c**).

Therefore, the behavior of the smaller pores interacting with each other seems to be driven by a competition between Ostwald ripening and splitting leading to an increased sphericity. However, as the total number of pores decreases at least for s-Ag or before $t_{crit}$ for s-Ag/Cu, it is likely that as expected, Ostwald ripening involves the larger pores (LP).

**Figure 9** shows an example of the time evolution of LP during aging for both s-Ag (**Figure 9a** and **Figure 9c**) and s-Ag/Cu (**Figure 9b** and **Figure 9d**) specimens respectively. As expected, the behavior of the larger pores LP is totally different from that observed for the smaller pores SP: their evolution is the same as that of the average behavior reported in **Figure 5** i.e. their volume increases following the Ostwald ripening mechanism except for s-Ag/Cu specimens when time reaches $t_{crit}$. After $t_{crit}$, their volume / radii begin to decrease (**Figure 9b** and **Figure 9d**).

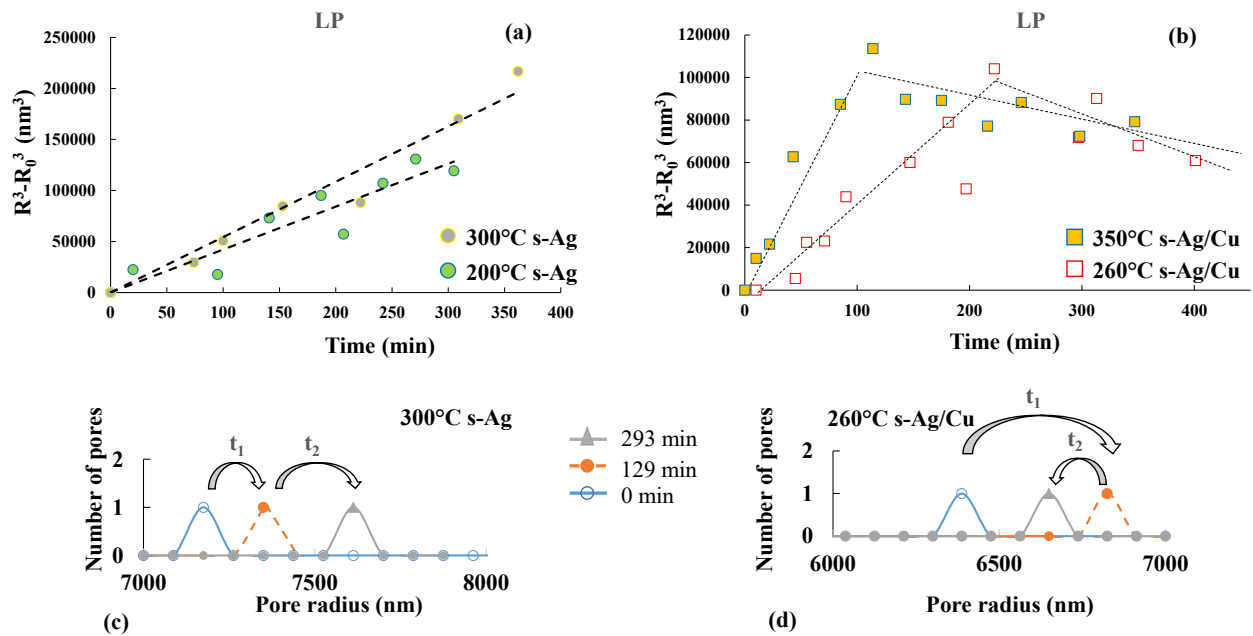

**Figure 9.** Evolution of LP with aging time ($t_2>t_1$) a) Ostwald ripening evolution of LP in s-Ag, b) Ostwald ripening of LP in s-Ag/Cu, c) in s-Ag, LP diameter increases with time and d) in s-Ag/Cu, LP diameter first increases ($t_1<t_{crit}$) then decreases ($t_2> t_{crit}$)

At this point it can be emphasized that the volume of smaller pores SP, interacting with each other, remains constant during aging by continuous connecting/splitting events. The evolution of the volume of the larger pores LP is nearly identical to the average evolution measured for the entire specimen. As a result, the evolution of the entire porous structure of sintered Ag is mainly controlled by the evolution of very few large pores LP. Those larger pores, representing over 80 % of the total initial pore volume, grow at the expense of the smaller ones, leading to a decrease of the total number of pores. The presence of the interface with Cu increases the evolution kinetics up to a critical time, after which LP volume / diameter starts to decrease. While the deviation from the LSW theory has been reported within stresses materials, a deviation from the Ostwald ripening mechanism in the course of aging (i.e. after a certain $t_{crit}$) has not been clearly yet reported. [33-38]

## 2.4. The role of the interface vs aging

The presence of the Cu/Ag interface seems to play a critical role in the evolution of the porous structure of sintered Ag. The pore evolution has been monitored using a specific s-Ag/Cu specimen which was aged at 300 °C. The specimens (described as "mixed specimens" in the following) were designed as follows: after Cu was deposited on the Ag cylinders, the specimens' sides were trimmed using Focused Ion Beam to obtain the geometry seen in **Figure 10a**. This geometry allows to monitor pore evolution from areas either close or far from the Ag/Cu interface (**Figure 10b**). As seen in **Figure 10c**, the pores close to the center (red triangles in Figure 10c) evolve according to Ostwald ripening. On the other hand, the evolution of the pores closer to the interface (yellow squares in **Figure 10c**), is much faster at the beginning then after a $t_{crit}$, deviates from Ostwald ripening. For sake of comparison, the evolution for s-Ag has been plotted in this figure. As observed in **Figure 10c**, Ostwald ripening kinetics at 300°C measured in any location for the mixed specimen are faster than that of pure s-Ag. Whether it is a normal deviation or a real difference in evolution kinetics is still under investigation.

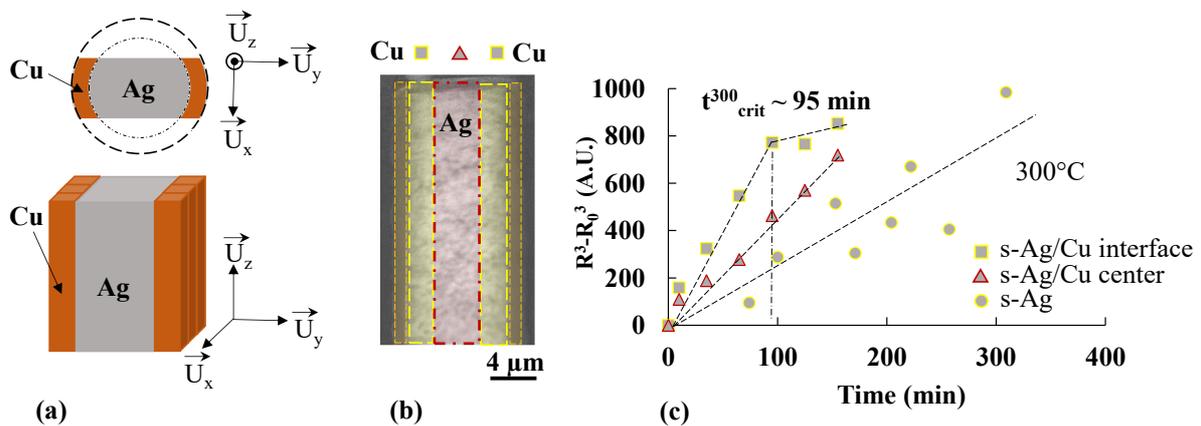

**Figure 10.** Evolution of the pores with aging depending on the distance from the interface (a) specific specimen design (b) different areas – close to the interface (yellow) – close to the center of Ag (red) (c) Pore radii evolution vs time (LSW representation) close to the Cu interface (grey/yellow square) or far from it (red triangle). s-Ag is given for reference (grey/yellow disc).

## 3. Discussion

### 3.1. Density

Densification of sintered Ag during aging is still an open question since both densification and non densification behavior were reported in the literature. In the previous studies, the density was either directly estimated using the pores surface fraction after image analyses or obtained by measuring the weight/volume ratio using high resolution scale and rulers. [5-9] Aging temperatures were in the range 125 °C to 350 °C for various exposure times. In the present study, while the spatial distribution of pores evolves, no densification was observed, even for temperatures well above the sintering temperature (240 °C). In other terms, sintering did not carry on after the initial processing step in the chosen aging conditions. This was observed for both pure sintered Ag (s-Ag) and copper coated Ag (s-Ag/Cu). In the case of s-Ag/Cu, this is even more surprising since thermal compressive stresses develop within Ag, resulting from the difference of thermal expansion coefficients between Cu ($19.0 \times 10^{-6}\,\text{K}^{-1}$) and Ag ($17.2 \times 10^{-6}\,\text{K}^{-1}$). These thermal stresses are higher than the sintering stress (10 MPa) (see Table 3 in section 3.3.). While densification cannot be ruled out for longer aging time, it has not happened during the present experiments although i) Ag self-diffusion coefficients at 350 °C are two orders of magnitude higher than at 240 °C and ii) aging time (420 min) is around 140 times longer than the usual sintering time (3 min).[44,45] Studies reporting densification with aging are often based on the evolution of the pore surface fraction. As previously mentioned (see **Figure 2**), surface analysis can be tricky/biased when dealing with complex 3D features. 3D imaging reveals that the pore diameters follow a binary distribution: most of the pores, representing over 99% of the total pore number, have a sub-micrometer diameter (SP). Only very few pores exhibit a diameter one order of magnitude higher (LP) (**Figure 1**). Actually, LP, while being very few, have a massive role within the pore fraction since they represent over 80 % of the overall pore volume fraction. Comparing data obtained after surface analysis (2D) to volume analysis may explain partially the discrepancy between the different studies found in the literature: As seen in the example shown in Figure 2b, 92 pores are counted in 2D while in reality, the volume from which this surface was extracted, contains only 13 pores. Among those 13 pores, one is very large, and its shape is very complex. As a result, surface analysis reveals many intercepts between this large pore and the surface. In the 2D analysis, each intercept is counted as a single pore (volumic connectivity cannot be considered since it is invisible) for which a diameter can be estimated. Interestingly, the average diameter $D_{2D}$ of those "independent pores" calculated using surface analysis, corresponds to pores belonging to the SP group.[5] As a result, surface analysis fails to reveal pores from LP. The shift towards larger radii values reported to occur during aging is not observed in the present study for which the SP average radius remains constant. In contrast, LP average radius changes as Ostwald ripening occurs. As a result, the shift reported to occur after surface analysis considers, at least partially, the growth of LP pores intercepting the surface. However, even for these two approaches, the result is expected to lead to the same conclusion: fewer pores combined with a volume increase of the larger ones lead to a constant density with time. What seems important here, is that the overall behavior is driven by the growth of a few larger pores (LP). As a result, the reliability of the system involving sintered Ag depends mostly on LP. It seems therefore very important to try to develop a material with a homogenous porous structure and, if possible, with the smallest possible pores. Based on the finding of the present study, this could help preventing the massive connectivity between pores and thus limiting the risks of weak spots (from both mechanical and thermal aspects).

## 3.2. Pore evolution during aging

Ostwald ripening was reported to occur at constant volume fraction of particles. To the authors' knowledge, the effect of the particle fraction on the kinetics of the Ostwald ripening is not well established (see for example [46-48]). Anyway, in the present study, this potential issue can be ruled out here since the volume fraction of pores is very close from specimen to specimen (**Table 1**). In contrast, elastic strains, resulting from the lattice misfit due to the coherent interface between a matrix and its precipitates, were shown to influence Ostwald ripening kinetics for a given temperature. [33-38] In the case of a porous material, coherency stresses can obviously be ruled out for the change in the Ostwald kinetics. However, power modules being composed of several stacked layers of various materials, stresses within the Ag joint can result from the processing (sintering of pure Ag) and/or from the difference of thermal expansion between the materials during aging. Using a Dynamic Resonant Method, Milhet et al. estimated residual stresses in the range of 0.4 MPa to 1.6 MPa just after processing in sintered Ag (using free standing specimens).[45] Using the same approach, thermal stresses developing at room temperature after sintering an Ag film on a Cu substrate at 240 °C were found in the range of 25 – 35 MPa (that is an order of magnitude higher compared to processing stresses) depending on the Ag film density. Processing stresses being much smaller than thermal stresses, they can be neglected. Furthermore, thermal stresses were shown to relax rapidly after thermal exposure for few hours by creep (although no time was indicated).[45]

The samples used for tomography analysis are not planar but cylindrical, i.e. far from a realistic configuration found in a power module. While the processing stresses are expected to be identical (pins dedicated to the tomographic experiments are machined from the same type of specimens as those used in [45]), the thermal stresses cannot be simply obtained using the planar approach. In the following, we will show that both configurations return similar stress levels.

### 3.3. Estimation of thermal stresses in cylindrical specimens coated with Cu

In order to estimate the stresses in our specimens, an infinitely long core-shell cylindrical structure is considered. It is composed of an Ag core of initial inner radius $R_{Ag}$, a shear modulus $\mu_{Ag}$ = 18 GPa, Poisson's ratio $\nu_{Ag}$ = 0.35 and thermal expansion coefficient (CTE) $\alpha_{Ag}$ = 19 × $10^{-6}$ K$^{-1}$ embedded in a Cu shell of outer radius $R_{Cu}$, shear modulus $\mu_{Cu}$ = 48.5 GPa, Poisson's ratio $\nu_{Cu}$ = 0.34 and thermal expansion coefficient $\alpha_{Cu}$ = 17.2 × $10^{-6}$ K$^{-1}$.[8, 45, 49, 50] The properties for sintered silver take into account the average pore volume fraction measured for each specimen according to the density except for the CTE that corresponds to bulk Ag since it was reported that the porosity doesn't alter the CTE.[49] Due to the thermal coefficients mismatch between the core and the shell, an eigenstrain can be considered into the core phase when core/shell interface is assumed to be coherent, as $\varepsilon_{ij}^* = \varepsilon_* \delta_{ij}$, with $\varepsilon_* = (T - T_0)(\alpha_{Ag} - \alpha_{Cu})$, $\delta_{ij}$ the Kronecker delta, and, finally, $T$ and $T_0$ = 293K the actual and ambiant temperatures, respectively.[51] The thermal strain and stress fields developing into the structure due to this eigenstrain can be determined, to the first order in $\varepsilon_*$, into the framework of the linear and isotropic elasticity theory as follows.[51, 52] Using the cylindrical coordinate system ($r, \varphi, z$), the general form of the elastic displacement field writes:

$$u_r^{Ag}(r) = A_{Ag} r, \quad u_\varphi^{Ag} = 0, \quad u_z^{Ag}(z) = C_{Ag} z, \qquad (2)$$

in the Ag core and,

$$u_r^{Cu}(r) = A_{Cu} r + \frac{B_{Cu}}{r}, \quad u_\varphi^{Ag} = 0, \quad u_z^{Cu}(z) = C_{Cu} z, \qquad (3)$$

in the Cu shell, with $A_{Ag}$; $C_{Ag}$; $A_{Cu}$; $B_{Cu}$ and $C_{Cu}$ five constants to be determined with the help of the following boundary conditions. Indeed, assuming the stress ($\bar{\bar{T}}^P$) and strain ($\bar{\bar{\varepsilon}}^P$) tensors are determined in the core ($p = Ag$) from Equation (2) and in the shell ($p = Cu$) from Equation (3) by means of the classical laws of elasticity, the mechanical equilibrium of the structure leads to the following set of Equations at first order in $\varepsilon_*$:

$$\sigma_{rr}^{Cu}(R_{Cu}) = 0, \tag{4}$$

$$\sigma_{rr}^{Cu}(R_{Ag}) - \sigma_{rr}^{Ag} = 0, \tag{5}$$

$$(R_{Cu}^2 - R_{Ag}^2)\sigma_{ZZ}^{Cu} + R_{Ag}^2\sigma_{ZZ}^{Ag} = 0, \tag{6}$$

$\sigma_{rr}^{Ag}$, $\sigma_{ZZ}^{Ag}$ and $\sigma_{ZZ}^{Cu}$ being constant into their respective phases. Likewise, the continuity of displacement at the core-shell interface gives at the first order in $\varepsilon_*$:

$$u_r^{Ag}(R_{Ag}) + \varepsilon_* R_{Ag} = u_r^{Cu}(R_{Ag}), \tag{7}$$

$$u_z^{Ag}(z) + \varepsilon_* z = u_z^{Cu}(z). \tag{8}$$

From Equation (4), Equation (5), Equation (6), Equation (7) and Equation (8), the different constants have been analytically determined and the stress components into the core have been found to be:

$$\sigma_{rr}^{Ag} = \sigma_{\varphi\varphi}^{Ag} = 2(1 + \upsilon_{Ag})(1 + \upsilon_{Cu})\mu_{Ag}\mu_{Cu}\varepsilon_* \frac{\psi_{Ag}^1}{\psi_{Ag}^2}, \tag{9}$$

$$\sigma_{zz}^{Ag} = 2(1 + \upsilon_{Ag})(1 + \upsilon_{Cu})\mu_{Ag}\mu_{Cu}\varepsilon_* \frac{\psi_{Ag}^3}{\psi_{Ag}^4}, \tag{10}$$

with

$$\psi_{Ag}^1 = (R_{Cu}^2 - R_{Ag}^2)\big((\mu_{Ag} - \mu_{Cu})R_{Ag}^2 + \mu_{Cu}R_{Cu}^2\big), \tag{11}$$

$$\psi_{Ag}^2 = -\mu_{Ag}^2 R_{Ag}^2(1 + \upsilon_{Ag})\big(R_{Cu}^2 + R_{Ag}^2(1 - 2\upsilon_{Cu})\big) + \mu_{Cu}^2(R_{Ag}^2 - R_{Cu}^2)^2(-1 + 2\upsilon_{Ag})(1 + \upsilon_{Cu}) - \mu_{Ag}\mu_{Cu}(R_{Ag}^2 - R_{Cu}^2)(-R_{Cu}^2(1 + \upsilon_{Cu}) + R_{Ag}^2(-2 + \upsilon_{Ag} + \upsilon_{Cu} + 4\upsilon_{Ag}\upsilon_{Cu})), \tag{12}$$

$$\psi_{Ag}^3 = (R_{Cu}^2 - R_{Ag}^2)\big((\mu_{Ag} - \mu_{Cu})R_{Ag}^2 + (\mu_{Ag} + \mu_{Cu})R_{Cu}^2\big), \tag{13}$$

$$\psi_{Ag}^4 = -\mu_{Ag}^2 R_{Ag}^2(1 + \upsilon_{Ag})\big(R_{Cu}^2 + R_{Ag}^2(1 - 2\upsilon_{Cu})\big) + \mu_{Cu}^2(R_{Ag}^2 - R_{Cu}^2)^2(-1 + 2\upsilon_{Ag})(1 + \upsilon_{Cu}) - \mu_{Ag}\mu_{Cu}(R_{Ag}^2 - R_{Cu}^2)(-R_{Cu}^2(1 + \upsilon_{Cu}) + R_{Ag}^2(-2 + \upsilon_{Ag} + \upsilon_{Cu} + 4\upsilon_{Ag}\upsilon_{Cu})). \tag{14}$$

Equivalent formulae hold for the stress field lying into the shell phase. The two constant stress components $\sigma_{rr}^{Ag}$ and $\sigma_{zz}^{Ag}$ (radial and longitudinal) displayed in Equation (9) and Equation (10) are plotted in **Figure 11a** and **Figure 11b** versus the outer radius $R_{Cu}$ for different temperatures $T = 200\,°C,\ 260\,°C,\ 300\,°C$ and $350\,°C$, with $R_{Ag} = 4$ µm. It is observed that the compressive stresses within sintered Ag ($\sigma_{zz}^{Ag}$ and $\sigma_{rr}^{Ag}$) increase rapidly (in absolute value) with the Cu layer thickness.

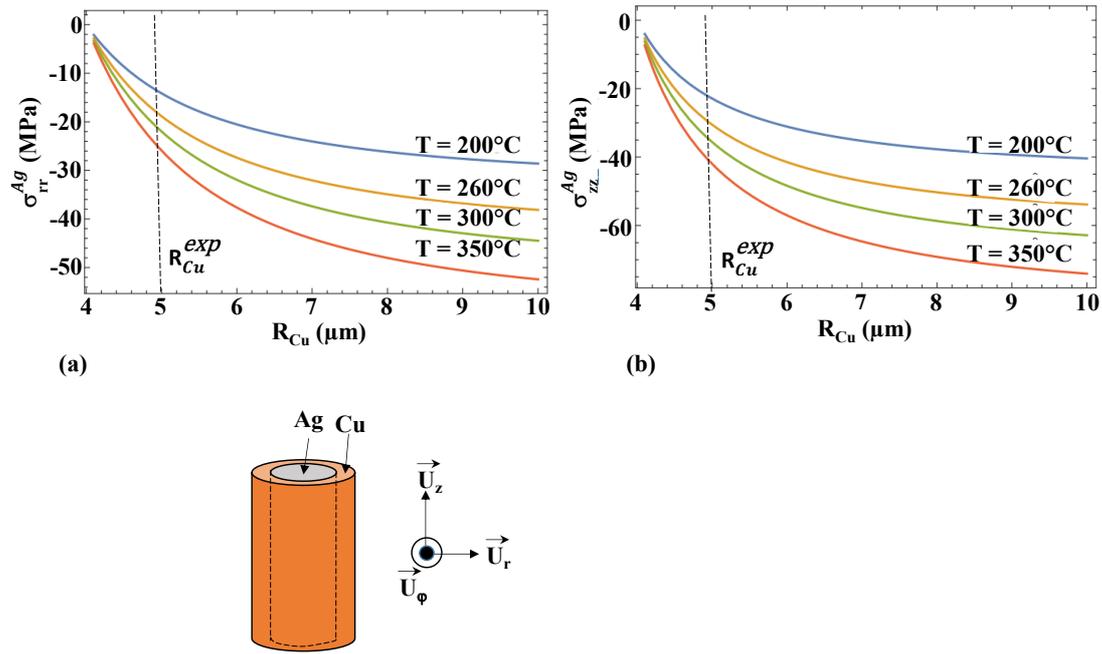

**Figure 11.** Radial ($\sigma_{rr}^{Ag}$) and longitudinal ($\sigma_{zz}^{Ag}$) thermal stresses within the Ag in s-Ag/Cu as a function of Cu thickness a) along $\mathbf{U_r}$ and b) along $\mathbf{U_z}$. The dashed lines represent our experimental conditions: an outer Cu radius $R_{Cu}^{exp}$ of 5µm, equivalent to a 1µm thick Cu shell around a 4µm Ag core

Interestingly, for an outer copper radius $R_{Cu}^{exp}$ of 5 µm, corresponding to 1 µm thick Cu layer around Ag (corresponding to our specimen configuration), the computed compressive stresses along the z axis $\sigma_{zz}^{Ag}$ ranges from -22 to -45 MPa depending on the temperature (200 °C to 350 °C).

A comparison of the compressive stresses estimated for the planar geometry (true to the real system – Cu thickness ranging from 0.5 mm to 1 mm and Ag thickness ranging from 20 μm to 30 μm) against those estimated for the cylindrical geometry is presented in **Table 3**. In the range of Ag diameters and Cu thickness chosen for the specimen dedicated to the tomography experiments, the stresses are very comparable. Indeed, the evolution of the porous microstructure presented in this manuscript can be seen as representative of the evolution expected in an Ag joint in a real system.

**Table 3.** Comparison between planar and cylindrical (this study) stresses

| Temperature [°C] | 200 | 260 | 300 | 350 |
|---|---|---|---|---|
| Planar [MPa] | -23 | -31 | -37 | -43 |
| Cylinder $\sigma_{zz}$ [MPa] | -22 | -30 | -36 | -45 |

Comparing the level of stresses within sintered Ag resulting from the thermal expansion difference with Cu to the mechanical properties as a function of the density of sintered silver at such temperatures is difficult since the data are scarce. Milhet et al. reported values at 125 °C for densities ranging from 6.5 g.cm$^{-3}$ up to bulk silver (10.5 g.cm$^{-3}$).[45] All the thermal stress values estimated in this section, calculated using properties true to the density, are well above yield stresses (YS) reported at 125 °C for the corresponding densities. YS at higher temperature (250 °C, 300°C and 350 °C) are expected to be even lower, creep of the Ag pin is likely to start rapidly once the target temperature is reached and help releasing the thermal stresses due to the presence of the Ag/Cu interface.

## 3.4. Pore evolution in the Ag/Cu system

The evolution of pores during aging is, at least until $t_{crit}$, driven by Ostwald ripening, i.e. by diffusion. The diffusion process can be investigated by analyzing the chemical potential of the diffusing species, i.e. vacancies in this case. It is well-know that this chemical potential depends on two terms, one proportional to the surface energy per unit surface γ times the surface curvature κ and a second term proportional to the elastic energy density.[39] This two-term dependence could thus explain the two different kinetic regimes of the pore evolution. At the beginning of aging of s-Ag/Cu samples, thermal stresses build up rapidly in Ag because of the Ag/Cu interface: the elastic energy term may be the driving force for faster Ostwald ripening kinetics. During this step (**Figure 12a**), the equilibrium shape of the pore is triggered by the balance between elastic energy (which helps developing surface roughness) and surface energy (symmetrical shape leaning towards a sphere).[40-42] This results both in a rapid increase of the larger pore volume along with a deviation from the natural tendency towards spheroidization. Thermal stresses in sintered Ag were shown to relax by creep during thermal aging.[45] In the present study, the stress within sintered Ag is therefore expected to decrease progressively down to 0 MPa leading to full relaxation. During the relaxation process, the stress reaches a threshold value below which the surface energy driven term eventually takes over the elastic one in the chemical potential. This assumption is well supported by concomitant evolution towards spheroidization of the larger pores and the deviation from Ostwald ripening after $t_{crit}$ (see example for s-Ag/Cu at 260 °C in **Figure 12a**). This diffusion mechanism aims to minimize the total energy in the cylinder (**Figure 12b**): the increasing number of pores can be explained in the framework of the stress-driven morphological instabilitiy of stressed solids. [39-43] Indeed, it is well-known that a wavy cylindrical-shaped pore with a finite wavelength λ can develop to reduce its surface energy. [53] This morphological evolution of the cylindrical pore has been found to be enhanced by stress which can lead to the final cut of the cylinder leading to a distribution of spherical pores.[53] This mechanism is assumed to take place here, where the very stringy nature of the larger pores can be optimized through the roughness development, the pore cutting and final increase of the number of spherical pores. As the temperature increases, this phenomenon is expected to be faster as thermal stresses build up and diffusion kinetics accelerate too.

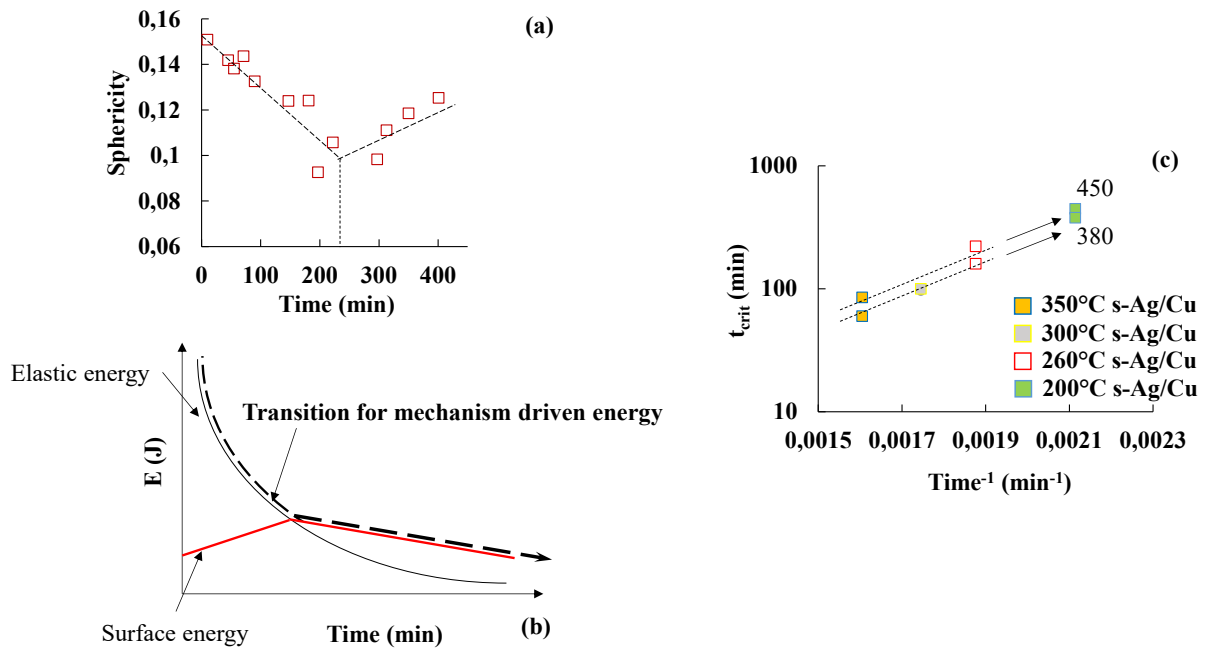

**Figure 12.** Evolution of the nanoporous structure vs aging time a) LP sphericity, b) Elastic and surface energy and c) critical time $t_{crit}$ before deviation from Ostwald ripening

Finally, the pore evolution observed when mixing both Ag/Cu interfaces and free Ag surfaces within a specimen leads to a mixed behavior. The thermal stress within Ag along the interface (along (Ox) and (Oz) axes – in-plane stresses) are expected to equal those from the cylindrical specimens (same interface, same temperature) close to the interface. Keeping in mind that in the very limited areas close to the corner of the interfaces and the free surfaces, the mechanical equilibrium of the structure leading to elastic relaxation may result in some stress variations, the following statements can be drawn. Since the thickness ratio between Cu and Ag is relatively large (close to 0.25), those stresses are expected to decrease when away from the Ag/Cu interface towards the specimen center. In this case, the Ag/Cu system can be seen as an Ag film on its planar Cu substrate. In contrast, along (Oy) axis, the thermal expansion of the system is not constrained, Ag and Cu can expand freely (**Figure 10a**). Away from those areas, close to the interfaces and free-surfaces, the stresses along this axis are close to zero. Overall, moving away from the interface, the pore structural evolution transit from an in-plane highly stressed configuration to a less stressed configuration. As a result, the aging behavior evolves from a faster Ostwald ripening followed by a deviation towards a more typical, linear Ostwald ripening behavior. Still, the stress state within Ag is not nil as it would for bulk Ag or an Ag/Cu with infinite Ag thickness. It is not clear at this stage if the stress state in the middle of Ag in the mixed Ag/Cu specimen tested at 300 °C would lead to a delayed deviation from Ostwald ripening or if the stress level is below a threshold value for which no deviation would occur. This is under investigation.

**Figure 12c** plots the evolution of $t_{crit}$ (y-axis in logarithmic scale), the critical time for the pore evolution within the s-Ag/Cu system to deviate from the ripening process as a function of 1/T. As observed in this plot, the critical times align well, indicating that $t_{crit}$ seems to follow a thermally activated phenomenon, i.e. diffusion following a classical Arhenius' law:

$$\frac{1}{t_{crit}} = \frac{1}{t_0} \cdot \exp\left(\frac{-Q_{tcrit}}{kT}\right) \tag{15}$$

with $t_0$ a constant, Q, the activation energy, k, the Boltzmann's constant and T, the temperature. Extrapolating at T= 200 °C gives a potential $t_{crit}^{200}$ within the range of 380 - 450 min. This is beyond the aging time performed during our study (aging at 200 °C was only performed for 350 min maximum) but the acceleration in pore evolution kinetics observed for the s-Ag/Cu specimen tend to be in favor of this behavior.

The rate coarsening constant K (as found in Equation 1) varies classically with temperature following a classical Arrhenius law:[28]

$$K \propto \exp\left(-\frac{Q}{kT}\right), \quad (16)$$

where Q is the activation energy for the diffusion, T the temperature and k the Boltzmann constant.

The activation energies $Q_{s-Ag}$, $Q_{Ag/Cu}$ and $Q_{tcrit}$, were estimated using Arrhenius plots for both K and $t_{crit}$. For $Q_{tcrit}$, only the experimental data obtained at 350°C, 300°C and 260°C were used. The activation energy for pure s-Ag $Q_{s-Ag} = 0.54 \pm 0.14$ meV while for the coated specimens Ag/Cu, $Q_{Ag/Cu} = 1.64 \pm 0.64$ meV. The activation energy estimated using $t_{crit}$ is $Q_{tcrit} = 1.77 \pm 0.32$ meV. Very interestingly, the activation energies $Q_{Ag/Cu}$ and $Q_{tcrit}$ are of the same order of magnitude, yet larger than the activation energy estimated in the case of the stress-free material. The mechanism responsible for the acceleration of the Ostwald ripening induced by the presence of the Cu coating (i.e. thermal stresses) seems therefore in close connection with the mechanism linked to $t_{crit}$. The increase of the activation energy between a stress-free state and a stressed material was already reported in the literature but is beyond the scope of this study.[54, 55] The deviation was not observed for the pores located close to the middle of the mixed specimen aged at 300 °C. In this latter case, the thermal stresses are lower than those estimated closer to the interface, but the ripening kinetics are accelerated compared to pure Ag. This could be an evidence of a different mechanism occuring during the evolution, closely linked to the level of stress within the material. In this context, as observed for the evolution of the pores in the middle of the mixed specimen tested at 300 °C, there is a possibility that the thermal stress developed within s-Ag/Cu at 200 °C remains sufficiently low such that the kinetics of the pores and thus the overall evolution of the material are controlled only by both stress and surfaces. This is currently under investigation. Anyway, it can be finally stated that all the processes involved in the porous structure evolution are closely related to thermal stresses and diffusion.

## 4. Conclusion

The evolution during aging in temperature of the porous structure of sintered Ag specimens with or without an Ag/Cu interface was investigated from 200 °C to 350 °C using in-situ X-ray nanotomography (BL 6.2.c – SLAC-SSRL, Menlo Park, CA, USA). With initial densities being very close to each other for each specimen, it was observed that:

The nanoporous structure consist of two clearly distinct pore populations: the smaller pores (SP group) represent over 80% of the total pore fraction number while very few bigger pores (LP group), with a diameter two orders of magnitude larger, represent over 80% of the total pore volume. The connectivity within the larger pores structure is very high, a key parameter that is only accessible through 3D nanotomography.

No densification was observed even for the higher aging temperatures.

The most significant structural evolutions are led by the larger pores: the higher the temperature, the faster the evolution. This evolution is driven by diffusion (Ostwald ripening), at least for a certain period of time depending on the specimen configuration (bulk or with an interface).

For a given temperature, Ostwald ripening kinetics are accelerated for sintered Ag containing an Ag/Cu interface compared to pure sintered Ag. This behavior results from the thermal stresses, arising from the thermal coefficient mismatch between Cu and Ag, and building up in Ag.

For sintered Ag containing an Ag/Cu interface, a deviation from Ostwald ripening is observed when a critical time, temperature dependant, is reached. This first observation of deviation from Ostwald ripening over a critical time is associated with the competition between stress relaxation and surface energy minimization.

## 5. Experimental Section

Specimen preparation for tomography was performed as described in [19]. Basically, Ag bulk films, as large as 3 x 1 x 0.5 cm$^3$, were produced using a Heraeus® silver micron paste with an average particle diameter of 4 µm following the processing route described in [45]. After a step during which each printed layer is dried under vacuum (controlled by DSC), final sintering is performed at 240 °C under 10 MPa uniaxial pressure for 3 min. This alternative approach to the classical two step procedure (drying at 150°C for 10 minutes followed by sintering at 240°C under 10 MPa for 3 minutes) leads to specimens with a porous structure similar to those of real joints.[45]

Transmission X-ray images were collected using the full-field transmission X-Rays microscope (TXM) at beamline 6-2c of Stanford Synchrotron Radiation Lightsource at the SLAC National Accelerator Laboratory.[19] The X-rays from a 56 pole 0.9 Tesla wiggler pass through several mirrors and are then focused to a spot of a few hundred microns, which acts as the virtual source for the microscope. The monochromator installed at this beamline is a liquid-nitrogen-cooled double-Si (111)-crystal system. The monochromator provides a quasi-monochromatic illumination over a 2.1– 17.0 keV range for the optics downstream. A mirror pitch feedback system is installed to monitor the micron-level beam movements and adjust the toroidal mirror in real-time for stabilizing the beam. The TXM, designed to work over an energy range from 5 keV to 14 keV, utilizes a capillary condenser to focus the beam to a spot of a few tens of microns. A Fresnel zone plate with 200 μm diameter and 30 nm outermost zone width is employed as the objective lens for a magnification of around 50 (depending on the energy of the incident X-rays). An optical objective lens is placed downstream from the scintillator crystal for a 10x optical magnification. A 2048 by 2048 pixel CCD is optically coupled with the scintillator crystal for image acquisition. The nominal spatial resolution of this system is approximately 30 nm as demonstrated using a standard resolution target. The TXM used in this study provides multiple contrast modalities, including absorption contrast and phase contrast. In the present study, we focused on the absorption contrast mode for two reasons: 1) the samples are made of high Z metals with sufficient absorption, and 2) we use the energy-dependence of the absorption contrast to enhance the sensitivity to the elemental compositions. More specifically, absorption-contrast tomography above and below the absorption edge of Cu, at 8962 eV and 8995 eV, respectively, is used to facilitate the segmentation of different sample components with good fidelity.

Data were collected from Ag micro-pillars whose geometry was optimzed for the TXM experiments (diameter ranging from 6 to 8 microns and a length ranging from 15 to 25 μm). Micro Pillars were prepared using the Nanobuilder software in a Ga-based ion beam Helios 600i FIB-SEM from FEI (now Thermo-Fischer) company (**Figure 13a**). It consisted of concentric rough milling at 9 nA beam intensity, followed by finer beam shaping at 2.5 nA. The last 1 to 2 microns in diameter were removed using a 700 pA beam to reach the final surface finish and the final dimensions. The ion beam was maintained parallel to the main axis of the pillar all the time, generating a 1-2° taper along the pillar due to the Gaussian spread of the ion beam.[56] Operating voltage was kept at 30 kV for all the processing steps.

Half of the specimens were pure sintered Ag (**Figure 13b**) while the other half were coated with 1 μm copper (Cu) to simulate an interface closely mimicking the one found in real systems (**Figure 13c**). The pillar being roughly cylindrical, Cu was electro-deposited using an acidic copper sulfate bath. The electrodeposition bath is composed of 0,5 mol.L$^{-1}$ of CuSO$_4$,5H$_2$O and 1 mol.L$^{-1}$ of Na$_2$SO$_4$,10H$_2$O with a pH adjusted between 1 and 1.5 by addition of H$_2$SO$_4$ solution at 2 mol.L$^{-1}$. Copper electrodeposition was performed using a conventional 3-electrode cell containing 400 mL of stirred solution at room temperature. A cylindrical Pt counter electrode and a calomel reference electrode (SCE) combined with a Luggin capillary were connected to a Modulab Solartron potentiostat driven by the XM studio software. Copper electrodeposits were obtained in galvanostatic mode at an applied cathodic current of 12 mA during 50 s. Previous analyses were performed to optimize the deposition conditions permitting to reach a 1 µm thick deposit. The deposition potential rapidly reached a constant of about -230 mV/SCE.

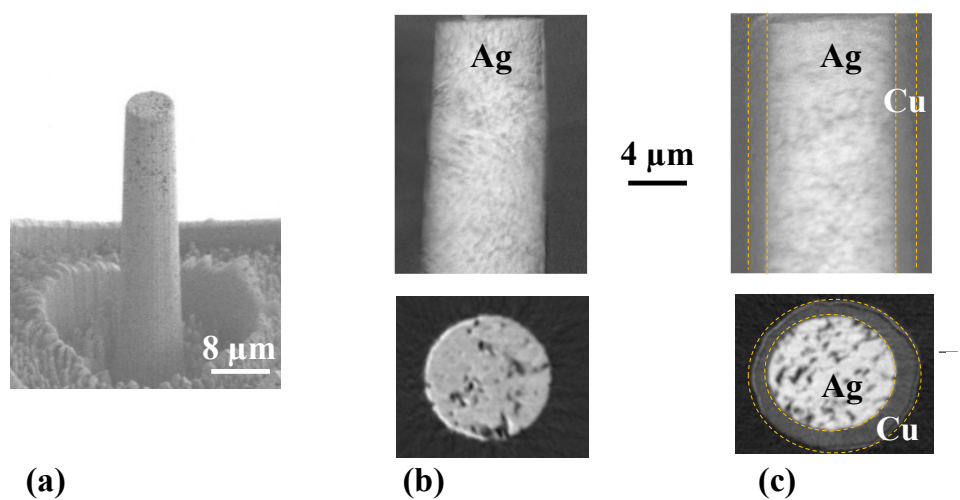

**Figure 13** a) SEM image of a pillar after FIB, b) sintered Ag and c) sintered Ag coated with 1 µm Cu viewed by in X-rays Computed Tomography

The samples were heated using an in-house developed heater inserted into the transmission X-ray microscope for in-situ control of the temperature at the sample position.[19] A cylindrical shaped copper heater shell is heated by two cartridge heaters with two thermocouples inserted at two different locations of the copper shell. A third thermocouple is used at the sample position
30

for calibration. This compact heater is capable of heating the sample up to ~ 600 °C at approximately 2 °C.s$^{-1}$ and the temperature variation within the heater is less than 1 °C. Aging temperatures were set to 200 °C, 260 °C, 300 °C and 350 °C respectively for up to 400 minutes, one specimen at a time, in isothermal conditions. The temperature range was chosen to give an insight of the pore evolution during aging in-service and to closely monitor the role of a Cu/Ag interface.

The imaging conditions were chosen to give the best of both worlds: high resolution and limited pore evolution between two subsequent scans. This was achieved by recording 180 projections over a rotation of 180° (1 projection per degree). A reference image (with no specimen in the field of view) was systematically recorded before and after each scan. Overall, the total acquisition time per scan is approximately 10 minutes. This is assumed to be fast enough to monitor the diffusion driven evolution in Ag (solid state) in the chosen temperature range (200 °C – 350 °C). No scan was recorded within the first 10 minutes after the initial heating at the target temperature **i)** to allow temperature stabilization and **ii)** help preventing artifact that may result from image drift because of the thermal expansion of the specimen's mounting rig. Ultimately, the chosen imaging condition is a tradeoff between the quality of the 3D tomogram and the temporal resolution, which needs to be carefully considered based on the targetted imaging performance.

The processing of tomographic volume of each sample was carried out using TXM Wizard® in three steps i) recovery of the 180 projections obtained directly after the thermal aging tests ii) automatic iterative alignment based on a phase correlation algorithm of the projections and iii) tomographic reconstruction of the aligned projections. The final format of the volumes obtained after the treatments with TXM Wizard is TIFF (Tagged Image File Format). [57, 58]

The reconstructed volumes were subsequently pre-processed using median filters available in AVIZO® 9.2 to enhance their quality while preserving the contours of the volume's constituents

After filtering, segmentation of microstructural constituents was carried out using region growing methods with AVIZO® 9.2. The estimation of the error generated by the segmentation procedure is done by comparing the covariograms of the raw and segmented image (see table 4). The covariogram allows us to obtain characteristic length of an image. [59, 60] These quantities, sizes of inclusions, characteristic distances, distance between pores… correspond to the quantities of interest for our study.

**Table 4:** Estimation of the error generated by the segmentation procedure done by comparing the covariograms of the raw and segmented image. [59, 60]

| Correlation length (Voxel) | BIN | RAW | Error (%) | Error (Voxel) |
| --- | --- | --- | --- | --- |
| X | 2.10E+01 | 2.00E+01 | 5.00% | 1.00E+00 |
| Y | 2.10E+01 | 2.10E+01 | 0.00% | 0.00E+00 |
| Z | 1.60E+01 | 1.60E+01 | 0.00% | 0.00E+00 |
|  |  | **average =** | **1.67%** | **3.33E-01** |

The 3D stacks were then analyzed using several algorithms based on mathematical morphology methods. [59, 60] These methods allow to separate the different pores using the skeletal calculation and a connectivity table for each voxel. [62, 63] This approach allows one to obtain the total number of pores, the volume of each pore, their centers of gravity and the distance between nearest neighbor pore surfaces (i.e. ligaments).

## 6. Acknowledgments

Use of the Stanford Synchrotron Radiation Lightsource, SLAC National Accelerator Laboratory, is supported by the U.S. Department of Energy, Office of Science, Office of Basic Energy Sciences under Contract No. DE-AC02-76SF00515. Computations have been performed on the supercomputer facilities at the Mésocentre de calcul SPIN hosted by the Université de Poitiers. FIB milling of pillars was performed within the framework of the French METSA network. This work pertains to the French Government program "Investissements d'Avenir" (EUR INTREE, reference ANR-18-EURE-0010).

The authors would also like to thank Dr Y.J. Liu and Dr. J. Nelson Weker (SLAC SSRL, Menlo Park, CA, USA) for their precious help during the experiments and helpflul comments. Finally, the authors would also like to thank Dr. L. Pizzagalli (Institut Pprime UPR 3346 CNRS, Poitiers, France) for fruitful comments on the manuscript.

Table of Content

Computed X-Ray Nano-Tomography is used to monitor the evolution of the porous structure of sintered silver coated with copper. While the pore evolution follows Ostwald ripening, the presence of thermal stresses near the interfaces promotes faster growth kinetics until a deviation from Ostwald ripening occurs. This behavior is associated with the competition between thermal stresses relaxation and surface energy minization.

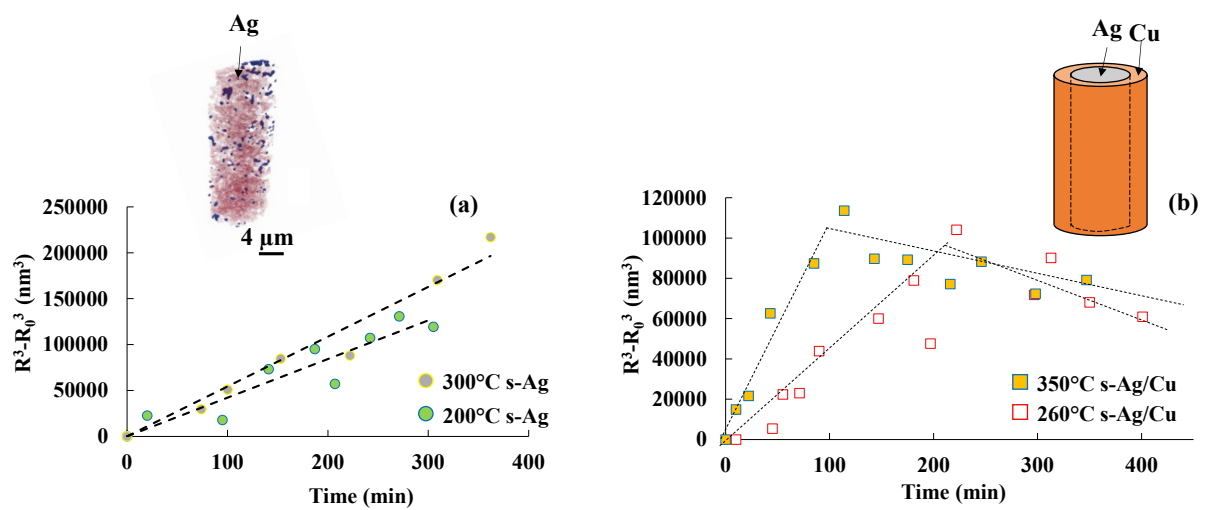

Authors Biographies

**Marc Legros** is Director of Research at CEMES-CNRS in Toulouse, France. He is focusing on the physics of plasticity through in situ Transmission Electron Microscopy (TEM) deformation of crystalline materials. Those can be structural (High Entropy Alloys, titanium or aluminum alloys) or oriented to energy conversion or harvesting (Al or Cu interconnects, silicon, uranium dioxyde). Among his metallurgy group at CEMES, he recently developed specific methods to quantify grain boundary processes in small-grained metals and alloys, where dislocation-based plasticity is shut down. ML is the recipient of the Bessel Awards, Constellium Prize and Silver medal of CNRS

**Juan Creus** is Professor in material sciences at La Rochelle University, France since 1998. . Research activities are focused on the relationship between coatings process, metallurgical states and functional properties. The research is linked on the influence of the incorporation of light elements like hydrogen, oxygen and carbon on the metallurgical state of electrodeposited coatings and on the embrittlement mechanism. Columnar growth or nanopores nucleated during electrodeposition could favor the hydrogen absorption affecting the mechanical performance. Innovative materials, Reach compliant are characterized by studying the corrosion mechanism in saline or marine environment.

**Loïc Signor** has defended its PhD Thesis at the University of Poitiers in 2008. He is lecturer at ISAE-ENSMA since 2009 and member of the department of physics and mechanics of materials at Institut Pprime (UPR3346, CNRS/ISAE-ENSMA/Université de Poitiers). His research topics include the study of microstructure – properties relationship in metallic polycrystalline alloys, using modelling approaches based on crystal plasticity finite element simulation.

**Jérôme Colin** is Professor at the University of Poitiers, France. His research is focused on the study of the mechanical properties and the plasticity of strained nanostructures. More precisely, the different topics addressed from both theoretical (theory of elasticity) and numerical (finite elements, molecular dynamics simulations) point of view concern the formation of dislocations in buried quantum dots, the delamination and buckling of thin films deposited on substrates and the morphological evolution of stressed solids in connection with the Asaro-Tiller-Grinfeld instability.

**Xavier Milhet** is Associate Professor at the University of Poitiers. His research is focused on the relationship between mechanical properties and microscrosture mostly for metallic materials (Ni-based superalloys, magnetic materials, electronic materials). More precisely, microstructure evolutions induced after exposure of the materials to harsh environment (high temperature) are characterized focusing on the relevant scale, responsible for the modification of the properties (including dislocations, precipitates and pores).